\documentclass[aps,pra,twocolumn,superscriptaddress]{revtex4}
\usepackage{graphicx}

\begin{document}

\title{Many-body phenomena in QED-cavity arrays}

\author{A. Tomadin} \affiliation{Institute for Quantum Optics and
  Quantum Information of the Austrian Academy of Sciences, A-6020
  Innsbruck, Austria}

\author{Rosario Fazio} \affiliation{NEST, Scuola Normale Superiore
  $\&$ INFM-CNR, Piazza dei Cavalieri 7, I-56126 Pisa, Italy}

\begin{abstract}
  Coupled quantum electrodynamics (QED) cavities have been recently
  proposed as new systems to simulate a variety of equilibrium and
  non-equilibrium many-body phenomena. We present a brief review of
  their main properties together with a survey of the last
  developments of the field and some perspectives concerning their
  experimental realizations and possible new theoretical directions.
\end{abstract}

\maketitle

\section{Introduction}
\label{sec:introduction}

The ability to design and fabricate controllable many-body systems has
been realized to be a precious tool to explore the world of strongly
correlated systems. In several different physical phenomena, ranging
from high temperature superconductivity to heavy fermions or the
fractional quantum Hall effect, strong local electronic correlations
play a crucial role. The seemingly simplified models to describe these
correlations, the Hubbard model of high temperature superconductivity
for example, are extremely difficult to solve and over the years a
number of judicious analytical and numerical methods have been
developed.  Together with these more ``traditional'' methods, in
recent years it was proposed that strongly correlated systems could be
studied by means of quantum simulators~\cite{UQS}, {\it
  i.e.}~fabricated systems that can experimentally simulate the model
Hamiltonian underlying the non-trivial properties of the physical
systems under consideration. The advantages of this approach are
twofold. First of all it is possible to explore the properties of
strongly correlated model Hamiltonians also in those regions of the
phase diagram which are elusive to numerical and analytical
investigations. Secondly it allows to test to which extent the model
Hamiltonians under consideration are appropriate to treat the physical
systems that they are supposed to describe or whether additional
ingredients are necessary.

Quantum simulators have a long and successful story.  Probably the
first fabricated system to have these characteristics were Josephson
junction arrays~\cite{FZ2001}. The field boosted with the appearance
of cold atoms in optical lattices~\cite{LSADSS2007} which proved to be
excellent simulators of a large variety of strongly interacting Fermi
and Bose systems.  Topic of the present paper is to review the latest
developments of a newly born direction in the field of quantum
simulators based on arrays of
QED-cavities~\cite{HBP2006,GTCH2006,ASB2007}.  There is already a
comprehensive review on the subject~\cite{HBP2008review}, in this
paper we will focus on some of the recent developments.

Cavity-QED arrays offer the possibility to realize strongly correlated
states of light. They can operate at high temperatures (as compared to
Josephson arrays and optical lattices) and allow for single-site
addressing thus opening a way to access experimentally correlation
functions. Furthermore they might allow to explore a number of new
equilibrium and non-equilibrium quantum phase transitions.

This paper is organized as follows. In the next section we will
introduce cavity arrays. We define the model Hamiltonians that govern
their dynamics and the most relevant sources of dissipation.  In
Sec.~\ref{sec:phasediagram} we discuss the most important
characteristics of the equilibrium phase diagram and the nature of the
low-lying excitations. Section~\ref{sec:simulators} describes how
cavity arrays can operate as quantum simulators.
Section~\ref{sec:nonequilibrium} reviews some properties of these
systems out of equilibrium. We complete the presentation with the
conclusions where we will briefly discuss possible experimental
realizations and some perspectives in the field.

\section{The Models}
\label{sec:models}

\begin{figure}[t]
  \includegraphics[width=0.9\linewidth]{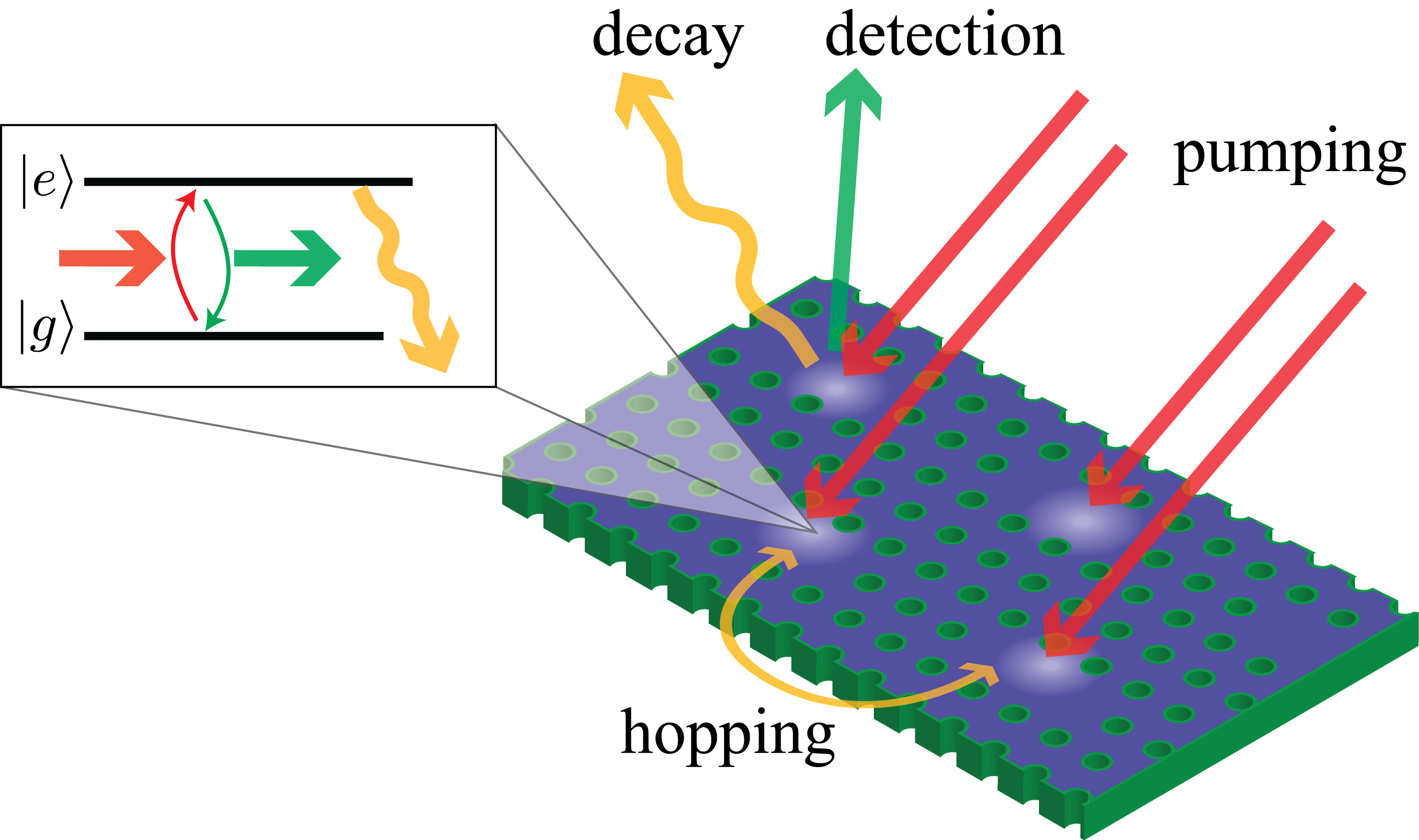}
  \caption{\label{fig:array} A sketch of a QED-cavity array. It
    consists of a regular arrangement of QED cavities. Neighboring
    cavities are coupled by photon hopping.  Nonlinearities in the
    cavities may produce an effective repulsion between the photons
    leading to an anharmonic spectrum.  The nonlinearity may be
    produced, {\it e.g.}~by a two-level system (depicted in the
    inset) coupled to the light resonating in the cavity and subjected
    to decay. Photons in the cavities have a finite lifetime therefore
    the cavities are pumped with an external coherent drive. }
\end{figure}

A sketch of a cavity array is illustrated in Fig.~\ref{fig:array}. It
consists of a regular arrangement of QED-cavities which are coupled by
the hopping of photons.  Light resonates in each cavity and interacts
with matter contained therein.  Experimental realization of cavity
arrays can be imagined in different ways. We will briefly discuss this
point at the end of the paper. Fig.~\ref{fig:array} illustrates the
case in which the array is realized using photonic crystals.  The
salient ingredient at the base of the rich physics of cavity arrays is
the interplay of two competing effects.  Light--matter interaction
inside the cavity leads to a (possibly strong) nonlinearity between
photons. On the other hand, photon hopping between neighboring
cavities favors delocalization thus competing with photon blockade.  A
model that describes a cavity array must then take into consideration
the interaction of the light and the matter within each cavity, the
coherent coupling between the cavities induced by the propagation of
the light, the leakage of photons out of the cavities, and the
dissipation and decoherence of the matter. It is also necessary to
include an external pump to populate the cavities and devise a method
to perform the measurement of the state of each cavity. We first
describe the terms that lead to the unitary evolution of the array and
afterwards the main sources of dissipation.

A single cavity confines several modes of the electromagnetic field
and each mode is quantized as an harmonic oscillator.  In the case
that a single mode with frequency $\omega_{\rm C}$ is considered, the
corresponding Hamiltonian is given by ${\cal H}_{\rm C} = \omega
a_{\ell}^{\dag}a_{\ell}$ where the operator $a_{\ell}$
($a_{\ell}^{\dag}$) annihilates (creates) a quantum of light in the
mode of the $\ell$-th cavity. If the cavities are sufficiently close
to allow for photon hopping, an additional kinetic term $-J (
a_{\ell'}^{\dag} a_{\ell} + {\rm H.c.})$ ($J$ is associated to the
tunneling rate) should be added to the Hamiltonian. These two terms
constitute the model for a cavity array in the absence of any
interaction with the matter field. In the presence of hopping the
Hamiltonian of the photons is still harmonic and can be diagonalized
by Fourier transform leading to the dispersion law of the photons in
the lattice.

The presence of atoms (quantum dots or Cooper pair boxes depending on
the implementation) inside each cavity leads to a strong effective
nonlinearity between photons. It is enough to think to the matter
field as a few-level system coupled to cavity mode and possibly to
some external source.  The Hamiltonian for an array of cavities can be
written as on each cavity and the photon hopping term between
different cavities
\begin{equation} \label{eq:fullham}
  {\cal H} = \sum_{\ell} {\cal H}_{\ell}^{(0)} - J
  \sum_{\left<\ell, \ell' \right>} ( a^\dagger_{\ell} a_{\ell'}+
  \mbox{H.c.})~.
\end{equation}
The local contribution ${\cal H}_{\ell}^{(0)}$ describes the
light--matter interaction.  In the limit in which leakage of photons
is ignored, the Hamiltonian in Eq.~(\ref{eq:fullham}) has been treated
in the grand-canonical ensemble. In this case one should add a term
containing the chemical potential.  In the next section the phase
diagram will be presented in the grand-canonical ensemble.

Probably the simplest model to describe the interaction between light
and matter is the Jaynes-Cummings model in which one mode of the
cavity interacts with a two-level system~\cite{SK1993}. One photon can
be absorbed by the two-level system, that goes into the excited state,
and conversely a photon can be emitted if the two-level system flips
from the excited $|2\rangle$ to the ground state $|1\rangle$.  The
Jaynes-Cummings model reads
\begin{equation} \label{eq:jaynescummings}
  {\cal H}^{(0)}_{\ell,{\rm JC}} = \varepsilon \sigma^z_{\ell} + \omega a^\dagger_{\ell}
  a_{\ell} + g ( \sigma^+_{\ell} a_{\ell} + \sigma^-_{\ell}
  a^\dagger_{\ell} )~,
\end{equation}
$ \sigma_{\ell}^{\pm}$ are the raising/lowering operators for the
two-level system and $\varepsilon$ denotes the transition energy
between the two levels.  In the rotating frame with respect to the
uncoupled Hamiltonian the relevant quantity is the detuning $\Delta =
\omega - \varepsilon$.  The spectrum of Eq.~(\ref{eq:jaynescummings})
is anharmonic so that, effectively, the two-level system induces a
repulsion between the photons in the cavity.  The strong effective
nonlinearity between the photons turns the cavity into a turnstile
device, where only a photon can be present at the same time.
Intuitively, this can be understood as the fact that one photon in the
cavity strongly modifies the effective resonance frequency, inhibiting
the injection of a second photon. This phenomenon has been termed
``photon blockade''~\cite{ISWD1997}, after the Coulomb blockade effect
of electrons in mesoscopic structures. The entire cavity thus behaves
as an effective spin system, that emits strongly anti-bunched light.
The cavity array Hamiltonian with the Jaynes-Cummings term was first
used by Greentree {\it et al.}~in Ref.~\cite{GTCH2006} and by
Angelakis {\it et al.}~in Ref.~\cite{ASB2007}.  A similar model with
many two-level systems in each cavity has also been used
\cite{RF2007,NUTY2008,LL2008}, where each two-level system interacts
with the mode of the cavity in a sort of Dicke-Bose-Hubbard model. Na
{\it et al.}~in Ref.~\cite{NUTY2008} allowed for the number of
systems to fluctuate among the cavities and showed that if the cavity
is strongly red-detuned with respect to the transition in the
two-level systems, the photons experience an effective Kerr
nonlinearity. Koch and Le Hur~\cite{KL2009}, though, pointed out that
the substitution of the Jaynes-Cummings coupling with an effective
Kerr nonlinearity is not appropriate in general.

Since the quantum effects in the cavity may be visible if the coupling
exceeds the decay amplitude of the matter and of the light, there has
been a lot of effort in increasing the effective cavity
nonlinearity. Strong interaction of light is in general limited by the
absorption of the medium, because the optical nonlinearities are weak
in non-resonant processes, while the absorption plays a dominant role
in the resonant processes.  One route to access the strong-coupling
regime involves the usage of external classical sources that prepare
coherently the matter in a state with reduced absorption.
Electromagnetically Induced Transparency (EIT) can be used to achieve
interaction strengths that are one order of magnitude larger than what
is possible with an ensemble of two-level
systems~\cite{FIM2005}. Various schemes to obtain strong
nonlinearities have been
discussed~\cite{SI1996,ISWD1997,HP2007,BHP2008}.  In this spirit
Hartmann {\it et al.}~in Ref.~\cite{HBP2006} considered an ensemble
of four-level atoms in a cavity array defined, in the rotating frame,
by a local interaction of the form
\begin{eqnarray}
  {\cal H}^{(0)}_{\ell ,{\rm EIT} } & = & \delta S^{33}_{\ell} + \Delta S^{44}_{\ell}
  + \Omega ( S^{23}_{\ell} + S^{32}_{\ell} ) \\
  & + & g_{1} ( S^{13}_{\ell}  \, a^\dagger _{\ell} + S^{31}_{\ell} \, a_{\ell} )
  +g_{2} ( S^{24}_{\ell}  \, a^\dagger_{\ell}  + S^{42}_{\ell} \, a_{\ell} )~, \nonumber
\end{eqnarray}
having defined the global atomic raising and lowering operators
$S^{lm} = \sum_{j=1}^N |l\rangle_j \langle m | $, $\sigma^{lm}_{j} =
|l\rangle_j \langle m |$ are the atomic raising and lowering operators
($l \neq m$), or energy level populations ($l = m$) for the $j$th
atom.  The transition $|3\rangle_j \to |2\rangle_j$ is driven by a
classical coupling field with Rabi frequency $\Omega$; the cavity mode
couples the $|1\rangle_j \to |3\rangle_j$ and $|2\rangle_j \to
|4\rangle_j$ transitions with coupling constants $g_1$ and $g_2$; the
parameters $\delta$ and $\Delta$ account for the detunings of levels
$3$ and $4$, respectively.  They considered a setup similar to
Ref.~\cite{ISWD1997} and introduced three families of polaritons,
delocalized over the whole atomic ensemble, that diagonalize the
Hamiltonian (if the state $|4\rangle$ decouples from the dynamics and
the number $N$ of atoms is very large).  One of the polaritons
projects only on the lowest-lying metastable states $|1\rangle$ and
$|2\rangle$ and is thus the longest-lived.  The polaritons can be
considered independent bosonic particles if their number is much
smaller than the number of atoms inside each cavity~\cite{footnote}.
The longest-lived polariton decouples from the others and experiences
an energy shift due to the perturbative coupling to the level
$|4\rangle$, that induces an interaction of the Kerr form. In this
way, in the large $N$ limit, they were able to show that the cavity
model maps onto an effective Bose-Hubbard model~\cite{FWGF1989} for
polaritons
\begin{equation}\label{eq:bhmodel}
  {\cal H}_{\rm BH} = -J \sum_{\langle \ell \ell' \rangle} a_{\ell'}^{\dag} a_{\ell} + \frac{U}{2} 
  \sum_{\ell} a_{\ell}^{\dag} a_{\ell}^{\dag} a_{\ell} a_{\ell}~.
\end{equation}

Up to now we did not take into account the leakage of photons out of
the cavities and the decoherence and dissipation of the few-level
systems.  One possibility, in the case of slow decay, is to complement
the Hamiltonian with imaginary frequencies multiplied by the
projectors of the decaying states, that render the dynamics
non-Hermitian (see some application of this method in
Ref.~\cite{FIM2005}).  Alternatively the decay can be given in terms
of the density matrix $\rho$ of the cavity array, using the master
equation
\begin{equation}\label{eq:masterequation}
  \partial_{t}\rho(t) = -i [{\cal H}, \rho(t)] + {\cal L}[\rho(t)]~,
\end{equation}
where the Liouvillian ${\cal L}$ in the Lindblad form reads {\it
  e.g.}
\begin{equation}\label{eq:liouvillian}
  {\cal L}[\rho(t)] = \frac{\kappa}{2} \sum_{\ell} \left ( a_{\ell} \rho(t) a_{\ell}^{\dag} - a_{\ell}^{\dag} a_{\ell} 
    \rho(t) - \rho(t) a_{\ell}^{\dag} a_{\ell} \right )~.
\end{equation}
This term describes leakage of photons from all the cavities of the
array, with an equal rate $\kappa$.  A similar term can be used to
describe decoherence and dissipation of the matter, although these
processes can usually be neglected in the timescale of the photonic
decay \cite{HBP2008review}.

In order to have a measurable signal out of the cavity array it is
necessary to refill with photons the modes of the cavities, to
contrast the leakage.  If a coherent laser beam of frequency
$\omega_{\rm L}$ is coupled to the cavity, the Hamiltonian acquires a
term $\varepsilon(t)_L a_{\ell}^{\dag} + \varepsilon(t)_L^{\ast}
a_{\ell}$, where $\varepsilon(t)_L$ is proportional to the electric
field of the beam.  To eliminate the explicit time-dependence $e^{i
  \omega_{\rm L} t}$ from $\varepsilon(t)_L$ it is convenient to use a
rotating frame in which the strength of the pumping is constant and
the frequency $\omega_{\rm L}$ is subtracted from the energy of the
photons in the cavity mode.  The contribution of the drive to the
Hamiltonian reads
\begin{equation}
  {\cal H}_{\ell, \rm D} = - \omega_{\rm L} a_{\ell}^{\dag} a_{\ell} + \varepsilon_L a_{\ell}^{\dag} + \varepsilon^{\ast}_L a_{\ell}~.
\end{equation}

In the next section we will discuss the main features of the phase
diagram deriving from the Hamiltonian~(\ref{eq:fullham}).

\section{Phase diagram}
\label{sec:phasediagram}

\begin{figure}[t]
  \includegraphics[width=\linewidth]{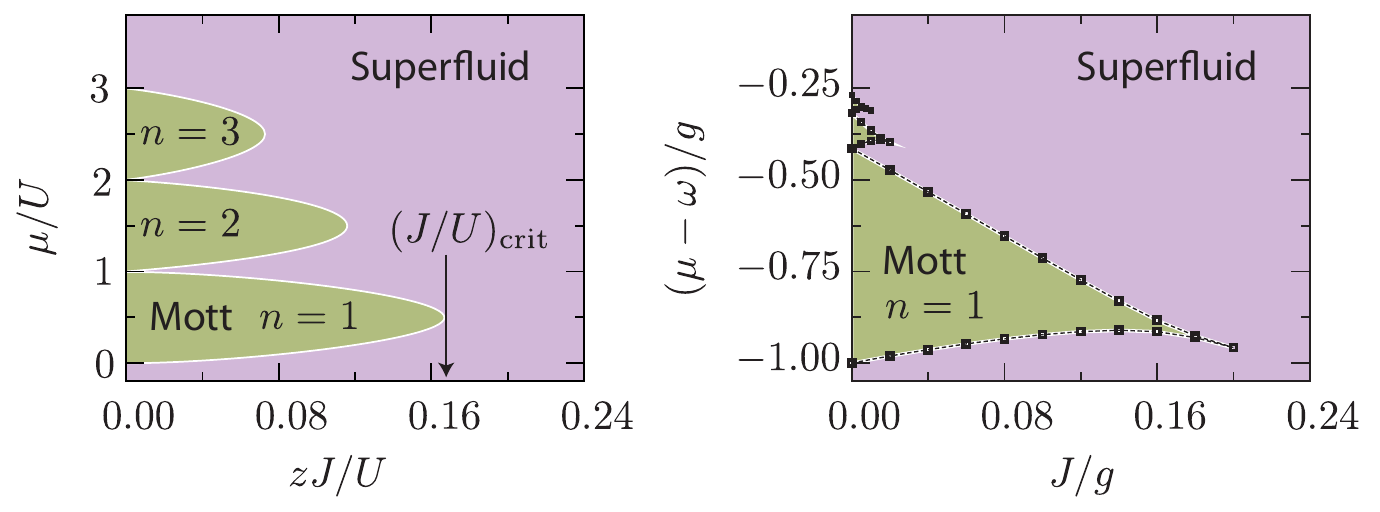}
  \caption{\label{fig:phasediagram} Left: sketch of the phase diagram
    of the BH model (\ref{eq:bhmodel}).  The green regions are the
    Mott lobes, $z$ is the coordination number of the lattice.  Each
    lobe extends up to a maximum critical ratio $(J/U)_{\rm crit}$.
    Right: the phase diagram of the cavity array in the case of a
    Jaynes-Cummings model, according to the DMRG computation presented
    in Ref.~\cite{RF2007}.  The green area corresponds to a Mott
    insulator of polaritons with average filling $n = 1$, $2$, and
    $3$. }
\end{figure}

The phase diagram of the Bose-Hubbard model defined in
Eq.~(\ref{eq:bhmodel}) has been studied for more than two decades (see
for example references in~\cite{LSADSS2007}). A qualitative
understanding of the zero-temperature phase diagram can be obtained by
considering the two limiting cases in which one of the two coupling
energies ($J$ or $U$) dominates. If the hopping dominates, the bosons
delocalize throughout the whole lattice.  In the limit of vanishing
interactions, the many-body ground state becomes simply an ideal
Bose-Einstein condensate where all the bosonic particles are in the
Bloch state of the lattice with vanishing quasimomentum, so that the
many-body wavefunction is written as $ \left (
  \frac{1}{\sqrt{L}}\sum_{\ell=1}^{L} a_{\ell}^{\dag} \right
)^{N}\,|{\rm vac}\rangle~, $ where $L$ is the total number of
cavities, $N$ is the total number of bosons and $|{\rm vac}\rangle $
is the vacuum state of the many-body Fock basis.  In contrast, if the
interaction dominates, each site has a well defined number of bosons
in the ground state. In order to put an extra boson on a given site,
one has to overcome an energy gap of the order of $U$.  In the Mott
lobes each cavity contains a definite number of bosons, the
fluctuations of the number operator vanish, and each particle is
localized in a given cavity. Each Mott lobe exists up to a maximum
hopping $(J/U)_{\rm crit}$ at which it is energetically convenient for
the particles to delocalize over the whole lattice (decreasing their
kinetic energy) rather than localize in a given site (decreasing their
mutual interaction). In the case of vanishing hopping, the Hamiltonian
is the sum of commuting local Hamiltonians and the ground state can be
written in the product form $\prod_{\ell}|\bar{n}\rangle_{\ell}$. When
the hopping is comparable with the onsite repulsion a quantum phase
transition takes place, separating a superfluid from a Mott insulator.
Although there are large quantitative differences between the case of
one-, two-, or three-dimensional lattices, the equilibrium phase
diagram of the model (at zero temperature, in the plane of the hopping
and the chemical potential) is similar in all cases and consists of a
superfluid phase surrounding a sequence of Mott-insulator lobes [see
Fig.~\ref{fig:phasediagram} (left panel)]. The $U(1)$ symmetry is
broken in the superfluid phase, where the kinetic energy dominates.

The phase diagram of the cavity array described by
Eq.~(\ref{eq:fullham}) resembles strong similarities with that of the
Bose-Hubbard model.  In Fig.~\ref{fig:phasediagram} (right panel) we
report for comparison the phase diagram obtained in Ref.~\cite{RF2007}
for a one-dimensional array.  Quantitative differences arise depending
on the model of the cavity.  For the sake of clarity we will ignore
these differences and try to highlight only those characteristics
which are common to cavity arrays.  The phase transition between the
Mott-insulator and the superfluid is a symmetry breaking phenomenon
that involves the conservation of the number of bosons $\sum_{\ell}
{n}_{\ell}$. The conserved quantity in the cavity model, take for
example the Jaynes-Cummings interaction, instead is the number of
polaritons, that do not have a definite quantum statistics. In cavity
arrays the detuning $\Delta$ can also be varied so that the phase
diagram becomes effectively three-dimensional \cite{KL2009}.  It turns
out that the detuning is a convenient experimental knob to tune the
system across the phase transition. Moreover, most importantly, to
change the detuning from negative to positive values drives the nature
of the excitations from quasilocalized excitons to polaritons to
weakly interacting photons~\cite{AHTL2008}.

A cavity array exhibits spectral properties similar to the BH models,
including gapped particle and hole bands in the Mott insulating phase
and Bogoliubov-type excitations in the superfluid phase.  The
single-particle excitation spectrum in the Mott phase has been
investigated also by Aichorn {\it et al.}~\cite{AHTL2008} using a
variational cluster approach, in which the self-energy of the infinite
lattice of cavities is approximated by a finite reference system.
Schmidt and Blatter~\cite{SB2009}, for the Jaynes-Cummings case, found
four modes corresponding to particle or hole excitations, versus the
two modes of the BH model. Pippan {\it et al.}~in Ref.~\cite{PEH2009}
studied the dynamical structure factor and the single-particle
spectrum using a quantum Monte Carlo simulation in one dimension. An
analysis of the nature of the excitations in small clusters was also
performed in Ref.~\cite{IOK2008}.  Paternostro {\it et al.}~in
Ref.~\cite{PAK2009} investigated the dynamical properties of an array
with many two-level systems in each cavity, beyond the elementary
excitation picture and found that the time-evolution of the array, in
the mean-field approximation, supports a soliton-like excitation.

Following the initial investigations~\cite{HBP2006,GTCH2006,ASB2007},
the existence of the Mott lobes in the cavity arrays has been verified
with several different methods. Makin {\it et al.}~\cite{MCTHG2008}
resorted to the exact diagonalization of small sets of cavities with
several topologies and to a cluster-mean-field approximation.  Schmidt
{\it et al.}~in Ref.~\cite{SB2009} proposed a strong-coupling theory
based on a linked-cluster expansion of the Green's functions at finite
temperature. This method includes quantum fluctuations beyond the
mean-field results, that are recovered instead in the random-phase
approximation. With these spectral methods the border of the Mott lobe
is signaled by the vanishing of the gap. A quantum Monte Carlo
analysis in two dimensions has been performed by Zhao {\it et al.}~in
Ref.~\cite{ZSU2008}.  The quantum Monte Carlo results for the phase
border compare quantitatively with the analytical
solution~\cite{SB2009} almost up to the tip of the Mott lobe. In one
dimension, the existence of the insulating Mott lobes has been
established in Ref.~\cite{RF2007} by means of a DMRG calculation. With
the same method, it has been shown in Ref.~\cite{RFS2008} that the
polariton fluctuations in a finite block of cavities yields a clear
signature of the two phases. The universality class of the transition
appears of the same type as the BH \cite{PEH2009,SB2009}.

The nature of the Mott-insulator for polaritons is different from the
BH case, in that the number of polaritons is fixed to an integer
value, but the fluctuations of the number of photons $\langle
{n}_{\ell} \rangle$ do not vanish. Similarly, in Ref.~\cite{PEH2009}
it was shown that the exciton and photon structure factors cannot be
used to characterize the Mott state. Although the photons are not the
genuine excitation of the system, it is desirable to characterize the
transition in terms of photonic observables since the emission of
light from the cavity can be measured in experiments. To measure the
polaritonic state using the projection on the atomic component
\cite{HBP2006,ASB2007} it is necessary to address the system with
further optical probes.

The presence of many two-level systems in each cavity and its effect
on the phase diagram has been analyzed in~\cite{NUTY2008,RF2007},
confirming that even with a moderate number of atoms in each cavity
there is a good quantitative agreement with the mapping onto the BH
performed in~\cite{HBP2006}. New features appear if the number of
atoms fluctuates from cavity to cavity. In this case a glassy phase
for polaritons is expected around the Mott lobes~\cite{RF2007}. Also
disorder in the distribution of the excitation and cavity
frequencies~\cite{NUTY2008}, and the effects of a finite
temperature~\cite{MCTHG2008,AHTL2008} have been considered.

Using more complex models of cavity arrays a two-component
Bose-Hubbard model~\cite{HBP2008} and superradiant Mott
phase~\cite{BHSS2009} have been shown to be realizable.

\section{Cavity arrays as quantum simulators}
\label{sec:simulators}

Building upon the models presented in Sec.~\ref{sec:models}, in
specific parameter ranges or with the addition of further optical
tools, it is possible to use the cavity array as an implementation of
other many-body models. In other words, the cavity array offers an
implementation of a ``quantum simulator'' for lattice models, in the
same spirit as the Josephson junctions arrays or the cold atoms in
optical lattices.  The first requirement for a quantum simulator is to
act as a calculator specifically tailored to the solution of the model
that it implements.  The knowledge on the model gained with the
quantum simulator can then be applied to all the other physical
systems described by the the same model. With respect to the
implementation of a quantum simulator in an optical lattice, the
cavity array may offer the advantage that each site of the array can
be addressed independently. The expectation values of the observables
could then be measured directly from the light emitted by one
cavity. Moreover, the cavity arrays can be implemented with several
different experimental systems and this may offer some advantages.  On
the other side, optical lattices seem unbeatable in terms of
scalability and absence of imperfections.  It should be said that at
present there are yet no experimental realizations of cavity arrays.
The hypothetical advantages of these quantum simulators need to be
tested against experimental realizations.

The earliest family of lattice many-body systems that have been
investigated in the context of cavity arrays is spin lattice systems.
The indication that a single cavity could act as an effective spin
system for the photons dates back to the recognition of the photon
blockade effect~\cite{ISWD1997}. Angelakis {\it et
  al.}~\cite{ASB2007} derived an effective XY Hamiltonian from the
Jaynes-Cummings model after decoupling the upper and the lower
polariton in the Mott polaritonic phase.  The realization of a ZZ
Ising coupling $\sigma_{\ell}^{z} \sigma_{\ell + 1}^{z}$ has also been
discussed \cite{LGGG2009} and the possibility to achieve both XY and
ZZ couplings has been put forward by Hartmann {\it et al.}~in
Ref.~\cite{HBP2007}.  The proposal relies on a cavity array containing
three-level atoms, in a $\Lambda$ configuration.  The two lowest-lying
states $|1\rangle$ and $|2\rangle$ are coupled to the excited state
$|3\rangle$ both via the quantized mode of the cavity and via external
classical sources. With an appropriate choice of the couplings and the
detunings, the dominant Raman transition between the two lowest-lying
states involves one laser and one cavity photon. This implies that the
emission and absorption of virtual photons in the cavity is
accompanied by a transition between the two lowest-lying states, that
become the effective $1/2$ spin in the cavity.  The coupling of the
virtual photons between neighboring cavities produces a XY model.
This effective spin can be coupled to an effective magnetic field in
the $z$ direction. With the same atomic spectrum but a different
configuration of the external sources, it is possible to implement a
ZZ interaction of the form $\sigma_{\ell}^{z} \sigma_{\ell + 1}^{z}$.
The XY and the ZZ couplings do not act at the same time, but are
combined in a unique effective Hamiltonian with the application of the
Suzuki-Trotter decomposition in which the lasers that produce either
interaction are periodically flashed one after the other.  Cho {\it et
  al.}~\cite{CAB2008} focused on the implementation of high-spin
Heisenberg models. Given the lack of analytical or numerical
informations that are available today on the phase diagram of
high-spin systems, the possibility to measure the phase diagram
directly from a quantum simulator is very interesting.  The scheme
includes an ensemble of three-level atoms in a configuration similar
to Ref.~\cite{HBP2007}, but relies on fixed external lasers.  The
model allows to simulate terms in the Hamiltonian of the form
${S}_{\ell}^{2}$, $({S}_{\ell}^{z})^{2}$, ${S}_{\ell}^{z}$,
${S}_{\ell}^{z} {S}_{\ell+ 1}^{z}$, ${S}_{\ell}^{x} {S}_{\ell+1}^{x}$,
${S}_{\ell}^{y} {S}_{\ell+1}^{y}$, and ${S}_{\ell}^{z}
{S}_{\ell+1}^{z}$.  Quite a few external sources are necessary to
control independently all the constants in the model, while the
magnitude of the spin is determined by the number of two-level systems
in the cavity.

Taking into account the polarization of light in the microcavities, Ji
{\it et al.}~in Ref.~\cite{JXL2007} put forward a model that is able
to exhibit ferromagnetism of the photonic circular polarization.  The
model considers an hexagonal photonic crystal with a square
superlattice of band gap cavities, doped with three-level systems.  In
the same system, it is also possible to access adiabatically the
antiferromagnetic phase, and an exotic super-counter-fluid phase in
which the total current of photons is zero as in the Mott phase, but
the currents with definite circular polarization are nonzero and
opposite.

The local addressability of cavity arrays has been exploited by Cho
{\it et al.}~in Ref.~\cite{CAB2008prl} to propose the simulation of
hard-core bosons with Abelian vector potentials. In particular this
method allows the simulation of the Laughlin wave function for the
fractional quantum Hall effect. The proposal assumes a two-dimensional
array of cavities, each containing a single three-level system.  To
address independently the two directions of the array, two modes of
each resonator are used, with the assumption that the frequency
difference between the two modes is much larger than the coupling
strength $g$. The simulation of the hard-core bosons follows from the
simulation of a spin system using the lowest-lying states of the
$\Lambda$ configuration, along the lines of Ref.~\cite{CAB2008}.  The
amplitude of the interaction between the spins on neighboring sites is
proportional to a position-dependent phase, that defines the gauge
potential.  This phase is controlled by the external laser sources
acting on the three-level system and the local addressability of the
cavities is invaluable to the precise implementation of this
phase. Moreover, by an adiabatic change of the laser phases, it is
possible to insert a flux quantum through the two-dimensional plane,
thus creating and moving a quasi-excitation in the system.

Cavity arrays can be used also to store and manipulate the resources
necessary to perform quantum computation.  The first study in this
direction has been made by Angelakis {\it et al.}~in
Ref.~\cite{ASYE2007} with the aim to realize quantum gates with
photonic-crystal waveguides. In this case the coupling between the
photons and the matter is not present in each cavity, but is used only
in some selected points of the quantum circuit to induce a phase gate
between the qubits.  The coherent control of the photonic transmission
through a cavity array has been considered by Hu {\it et al.}~in
Ref.~\cite{HZSS2007}, where it was shown, by a Green's function
technique, that the dispersion of the hybrid light--matter excitations
in the array can be controlled by acting on the population inversion
of the dopants. A procedure to generate indistinguishable single
photons or polarization-entangled photonic pairs, which are important
resources to the implementation of the quantum computation algorithms,
has been demonstrated in Ref.~\cite{NY2008} by Na {\it et
  al.}~resorting to the exact integration of a small number of
cavities.  The complete implementation of the Grover's search
algorithm has been proposed recently by Kyoseva {\it et
  al.}~\cite{KAC2009} together with a proposal to implement one-way
quantum computation with cavity arrays~\cite{KA2008}.

\section{Non-equilibrium behavior}
\label{sec:nonequilibrium}

\begin{figure}[t]
  \includegraphics[width=0.8\linewidth]{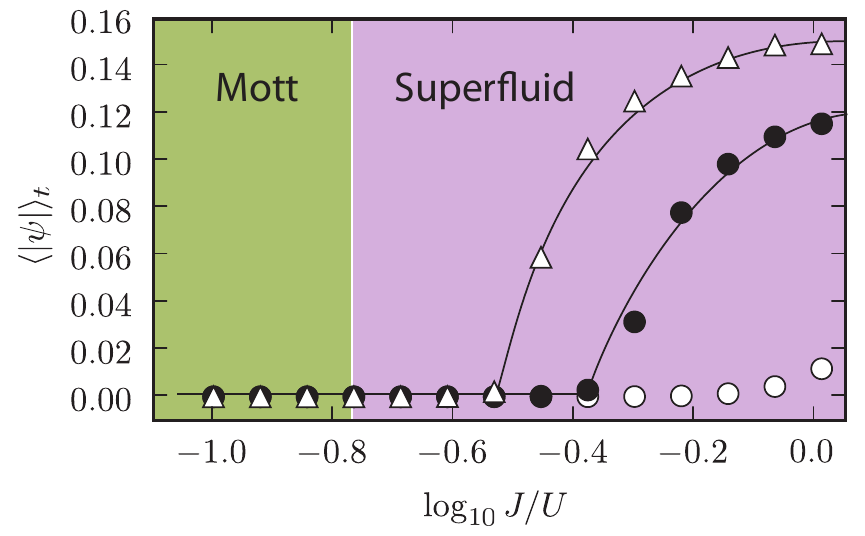}
  \caption{ \label{fig:nonequilibrium} Non-equilibrium signature of
    the quantum phase transition between the Mott-insulator and the
    superfluid state in the Bose-Hubbard model (\ref{eq:bhmodel}) with
    the leakage of photons described by (\ref{eq:liouvillian}),
    obtained by Tomadin {\it et al.}~in Ref.~\cite{TGFGCTI2009}.  The
    order parameter $\psi = \langle a \rangle$, averaged in an
    interval of time, is shown for $U/\kappa = 50$ (empty circles),
    $100$ (filled circles), and $200$ (empty triangles). The
    transition of the order parameter takes place together with a
    transition from antibunched to bunched light.  On decreasing the
    dissipation strength the onset at which the order parameter
    becomes non-zero approaches the equilibrium value (indicated as a
    vertical line). }
\end{figure}

In order to detect experimentally the different quantum phases
presented in Sec.~\ref{sec:phasediagram} one assumes that the mixed
light--matter excitations have a lifetime that is longer than the time
necessary to the measurement process.  Similarly, in order to perform
the quantum simulations of a ground state discussed in
Sec.~\ref{sec:simulators} it is necessary that the timescale of the
many-body dynamics is shorter than the decay time.  For example, the
ground state of a many-body system could be reached if the decay of
the polaritons is at least slower than the intercavity tunneling
\cite{NUTY2008}. As the decay times of the photons and of the matter
strongly depend on the specific implementation \cite{HBP2008review},
it is interesting to consider schemes that allow to measure the
signatures of the many-body physics without relying on a negligible
dissipation. It is then necessary to study the cavity array under
strong non-equilibrium conditions.

Carusotto {\it et al.}~in Ref.~\cite{CGTDCI2009} investigated a
system of photons in a circular array of cavities in the
Tonks-Girardeau limit.  All the cavities are driven by a single laser
beam and the far-field emission, that gives access to the occupation
of the single-particle states in the momentum representation, is
computed. The laser couples to the many-body wavefunctions supported
by the array depending on the total energy of the state. This yields a
spectroscopic analysis of a many-body system that is created by the
probe beam itself.  In the limit of very strong Kerr nonlinearity, the
photons become hard-core bosons and, being in one dimension, are
represented by an effective wave function for noninteracting fermions.
The classification of the states given by this mapping is used also to
characterize the absorption in the case of intermediate interaction
strength.

The quantum phase transition of the BH (see
Sec.~\ref{sec:phasediagram}), in an open cavity array, has been
investigated by Tomadin {\it et al.}~in Ref.~\cite{TGFGCTI2009} who
proposed to discriminate the Mott-insulator and the superfluid phases
analyzing the light emitted by the cavity array following a pulse that
creates a Mott state in the system.  The existence of the two phases
is clearly seen in the coherence properties of the emitted light (see
Fig.~\ref{fig:nonequilibrium}) for currently achievable values of the
ratio between the interaction and the dissipation.  The method relies
on the dynamical instability of the BH that follows a quantum quench
from the Mott to the superfluid phase.  The realization of the quantum
quench in the dissipative environment is realized with a careful
design of the initial pulse, that is substantially different from a
$\pi$-pulse, and it is shown that the detection of the phase border is
very robust to imperfections in the pumping. The time-evolution of the
system produces antibunched light if the ratio $J/U$ is smaller than
the critical value $(J/U)_{\rm crit}$ at the tip of the Mott lobe, and
the crossover from antibunched to bunched light clearly marks the
phase boundary.

\section{Conclusions}

As candidates for simulating strongly interacting models, coupled
cavities present new characteristics as compared to other successful
examples, like optical lattices or Josephson junction arrays: most
notably it is possible to access their local properties. In this paper
we briefly review their basic properties and some of the latest
developments in the field.
  
QED-cavities can be realized in a number of different
ways~\cite{wallraff04,hennessy,faraon}. This flexibility in the design
is a potential advantage to realize different local
nonlinearities. Moreover different systems may allow for different
measurement schemes as well. As already mentioned, at present there
are no experimental realizations of cavity arrays although it seems
that technological requirements are already at hand to fabricate a
small number of coupled cavities. In this respect it may be
interesting to explore properties of small clusters.

In this review we confined the discussion to coupled cavities. There
are other systems in which many-body physics with light can be
realized (see for example~\cite{chang,kiffner}). Due to space
limitations we did not touch these very interesting directions of
research.

\section*{Acknowledgments}

We acknowledge a very fruitful collaboration with Iacopo Carusotto,
Dario Gerace, Vittorio Giovannetti, Atac Imamoglu, Davide Rossini,
Giuseppe Santoro, and Hakan Tureci. We benefitted from many
discussions with Dimitris Angelakis, Michael Hartmann, and Martin
Plenio. The work was supported by EU project IP-Eurosqip.


\begin{thebibliography}{99}
\bibitem{footnote} The notation $a_{\ell}$ is used to denote exact
  (photons) and approximated (polaritons) bosonic particles.
\end{thebibliography}
\end{document}